\documentstyle[epsf,12pt]{article}
\newcommand {\beq}{\begin{equation}}
\newcommand {\eeq}{\end{equation}}
\newcommand {\beqa}{\begin{eqnarray}}
\newcommand {\eeqa}{\end{eqnarray}}
\newcommand {\n}{\nonumber \\}

\renewcommand{\theequation}{\thesection.\arabic{equation}}
\begin{document}
\setlength{\oddsidemargin}{0cm}
\setlength{\baselineskip}{7mm}

\begin{titlepage}
 \renewcommand{\thefootnote}{\fnsymbol{footnote}}
$\mbox{ }$
\begin{flushright}
\begin{tabular}{l}
KEK-TH-949\\
YITP-04-16\\
Mar. 2004
\end{tabular}
\end{flushright}

~~\\
~~\\
~~\\

\vspace*{0cm}
    \begin{Large}
       \vspace{2cm}
       \begin{center}
         {Correlators of Matrix Models \\
on Homogeneous Spaces}
\\
       \end{center}
    \end{Large}

  \vspace{1cm}

\begin{center}
           Yoshihisa K{\sc itazawa}$^{1),2)}$\footnote
           {
e-mail address : kitazawa@post.kek.jp}
           Yastoshi T{\sc akayama}$^{2)}$\footnote
           {
e-mail address : takaya@post.kek.jp}\\{\sc and}
           Dan T{\sc omino}$^{3)}$\footnote
           {
e-mail address : dant@yukawa.kyoto-u.ac.jp}

        $^{1)}$ {\it High Energy Accelerator Research Organization (KEK),}\\
               {\it Tsukuba, Ibaraki 305-0801, Japan} \\
        $^{2)}$ {\it Department of Particle and Nuclear Physics,}\\
                {\it The Graduate University for Advanced Studies,}\\

{\it Tsukuba, Ibaraki 305-0801, Japan}\\
        $^{3)}$ {\it Yukawa Institute for Theoretical Physics, Kyoto
University}\\
           {\it Kyoto 606-8502, Japan}
\end{center}

\vfill

\begin{abstract}
\noindent
We investigate the correlators of
$TrA_{\mu}A_{\nu}$ in matrix models on homogeneous spaces:
$S^2$ and $S^2\times S^2$. Their expectation value is a good order
parameter to measure the geometry
of the space on which non-commutative gauge theory is realized.
They also serve as the Wilson lines which carry the minimum momentum.
We develop an efficient procedure to calculate them
through 1PI diagrams.
We determine the large $N$ scaling behavior of the correlators.
The order parameter shows that fuzzy $S^2\times S^2$
acquires a 4 dimensional fractal structure
in contrast to fuzzy $S^2$.
We also find that the two point functions exhibit
logarithmic scaling violations.

\end{abstract}
\vfill
\end{titlepage}
\vfil\eject

\section{Introduction}
\setcounter{equation}{0}
One of the most important problems in string theory
is to understand the dynamics of D-branes.
The understanding of such a problem is likely to
lead to the ultimate formulation of string theory.
For example, old matrix models for non critical strings
are reinterpreted as matrix models of D-branes\cite{MGHV}\cite{KMS}.

Such a development reinforces our interest in
matrix models for critical strings\cite{BFSS}\cite{IKKT}.
In matrix models,
non-commutative solutions are identified as D-branes.
On such a solution,
non-commutative (NC) gauge theory is realized\cite{CDS}\cite{AIIKKT}\cite{Li}.
In string theory, NC gauge theory arises on D-branes with
$B_{\mu\nu}$ field\cite{SW}. More recently
the non-commutativity parameter
is identified with $g_s\alpha '$ where $g_s$ is the string
coupling constant and $\alpha '$ is the inverse
string tension\cite{MW}\cite{Vafa}. We therefore suspect that
non-commutativity plays a fundamental role in string theory\cite{Yoneya}.

In a non-perturbative investigation, it is desirable to
work with matrices of finite $N$. In such a setting,
branes are inevitably compact. The simplest such objects
in matrix models are fuzzy sphere and its higher dimensional
analogues\cite{Madore}\cite{KN}\cite{ZH}.
In order to obtain such a classical solution in matrix models
with finite $N$, we need to introduce Myers term and its
generalizations\cite{Myers}\cite{IKTW}\cite{Mathom}.
It serves as a nonperturbative formulation of NC gauge theories
on homogeneous spaces $G/H$.

We have investigated quantum corrections of NC gauge theory
on $G/H$\cite{fuzS2}$\sim$\cite{IT}.
We have identified the 't Hooft couplings and large $N$ scaling
behavior of these gauge theories by power counting arguments.
Recently a non-perturbative study has been performed in a bosonic
matrix model\cite{ABNN}.
The large $N$ scaling behavior of the observables
are found to be in good agreement with perturbative
predictions.

From string theory point of view, the Myers term corresponds
to the constant $H_{\mu\nu\rho}$ field.
With finite $N$, most of the solutions are metastable since
the single $S^2$ solution always minimizes the action at the classical level.
Nevertheless we expect that they are stabilized
in the large $N$ limit due to the suppression of the tunneling effect.
We have indeed found that the $H_{\mu\nu\rho}$ field vanishes in our large $N$ scaling analysis.
We therefore argue that various $G/H$ solutions approach non-commutative
$R^d$  with various dimensionality $d$ in the large $N$ limit 
realizing the formal classical solutions of IIB matrix model.
Our investigations concern quantum fluctuations around a particular
extremum of the effective action and the exploring the entire landscape
is beyond the scope of this paper.

Another possibility is to
contemplate that such a solution extremizes the quantum effective action
of IIB matrix model.
In fact we have found that a 4 dimensional solution of $S^2\times S^2$ type
extremizes the effective action at the two loop level\cite{fuzS2S2}\cite{IT}.
Since its effective action is $O(N)$ which is the same order
with the space-time volume, we may interpret this solution as a D3 brane
with a finite tension.
It is because the true minimum of the effective action
is argued to be $O(1)$ in IIB matrix model\cite{KNS}\cite{MNS}.

In the matrix model construction of NC gauge theory,
space-time and gauge field are unified into the identical
matrix degrees of freedom $A_{\mu}$ which are identified as the
coordinates semiclassically.
In this sense the gauge invariant operator $TrA_{\mu}A_{\nu}$
is a good order parameter to measure the shape and extension of the
Euclidean space-time on which NC gauge theory is realized.

In this paper, we investigate the vacuum expectation value
(one point function) of these operators in detail.
We develop an efficient procedure to calculate
them through 1PI diagrams. We perform explicit calculations
on $S^2$ and $S^2\times S^2$ in various matrix models
up to the two loop level.
We identify the large $N$ scaling behavior
of the correlators
to all orders by power counting arguments.
We find that the correlators on $S^2$ and $S^2\times S^2$ exhibit
different scaling behavior due to different
dimensionality. In particular $<TrA_{\mu}A_{\nu}>$ on $S^2\times S^2$
indicates a 4d fractal like structure of it
in contract to $S^2$ which remains to be semiclassical.

The organization of this paper is as follows.
In section 2, we investigate the effective action
of quantum fields at the one loop level in preparation to calculate
the correlators at the two loop level.
We discuss how gauge invariance is respected in NC gauge theory.
In section 3, we calculate the vacuum expectation value
of this operator on $S^2$ up to the two loop level.
We first evaluate it by the connected diagrams.
We then evaluate it by 1PI diagrams after
expanding $A_{\mu}$ around the quantum solution which extremizes
the effective action.
We find that the both methods produce the identical
answer as it can be argued on general grounds.
The order parameter is found to consist of
the Kronecker's $\delta$ function of 3d space in which $S^2$ extends.
In section 4, we calculate the corresponding order parameter on $S^2\times S^2$
by 1PI diagrams.
We find that it consists of the two independent
tensors. We interpret the new tensor as the
contribution from 4d fractals.
In section 5, we explain that these operators also serve
as the Wilson lines on homogeneous spaces\cite{IIKK}\cite{Gross}.
We compute the connected two point functions
of these Wilson lines.
We conclude in section 6 with discussions.

\section{Effective action}
\setcounter{equation}{0}

In this section, we investigate the one loop
self-energy in NC gauge theory on $S^2$.
This investigation is useful to compute
the gauge invariant correlators up to the
two loop level.
Since NC gauge theory is realized by matrix models
with finite $N$, the gauge invariance is
exactly maintained.
Our first concern in this section is
how the gauge invariance is reflected by
the one loop self-energy of gauge fields.

Let us consider IIB matrix model:
\beq
S_{IIB}  =  -{1\over g^2}Tr({1\over 4}[A_{\mu},A_{\nu}][A^{\mu},A^{\nu}]
+{1\over 2}\bar{\psi}\Gamma ^{\mu}[A_{\mu},\psi ]) .
\label{action}
\eeq
We expand $A_{\mu}$ around a background ($p_{\mu}$) which
represents a fuzzy $S^2$:
\beq
A_{\mu}=f(p_{\mu}+a_{\mu}).
\eeq
where $a_{\mu}$ denotes the gauge field on it.
$p_{\mu}$ can be identified with the angular momentum operators
in the spin $l$ representation where $N=n(2l+1)$
in the case of $n$ identical spheres.
$f$ is the scale factor of this background.

At the one loop level, the effective action
consists of the gauge sector (gauge field and ghost) and
the fermion sector contributions.
The both contributions must be BRS invariant
and the fermion sector contribution must be gauge invariant
in particular.
The fermionic contribution to the one loop effective action is
\beqa
&&-{1\over 2}Trlog(1+\Gamma\cdot \delta A{1\over \Gamma\cdot P})\n
&=&
-{1\over 2}Tr[\Gamma\cdot \delta A{1\over \Gamma\cdot P}]+
{1\over 4}
Tr[\Gamma\cdot \delta A{1\over \Gamma\cdot P}
\Gamma\cdot \delta A{1\over \Gamma\cdot P}]
+\cdots ,
\eeqa
where $\delta A_{\mu}X=[a_{\mu},X]$.
In this expression, the first and second terms represent the tadpole
and self-energy of gauge field respectively.

The gauge field self-energy (two point function)
in the above expression consists of
the planar and non-planar contributions:
\beqa
&&-{1\over 2}
<a^{\mu}|tr\Gamma_{\mu}{1\over \Gamma\cdot Q}\Gamma_{\nu}
{1\over \Gamma\cdot R}
|a^{\nu}>_p \n
&&+{1\over 2}
<a^{\mu}|tr\Gamma_{\mu}{1\over \Gamma\cdot Q}\Gamma_{\nu}
{1\over \Gamma\cdot R}
|a^{\nu}>_{np} ,
\label{fself}
\eeqa
where the symbol $tr$ evaluates the trace over spinor indices only and
\beqa
P_{\mu}Y_{j_1m_1}&\equiv& [p_{\mu},Y_{j_1m_1}],\n
Q_{\mu}Y_{j_2m_2}&\equiv& [p_{\mu},Y_{j_2m_2}],\n
R_{\mu}Y_{j_3m_3}&\equiv& [p_{\mu},Y_{j_3m_3}].
\eeqa
We have also introduced the following average:
\beqa
<a^{\mu}|X|a^{\nu}>_p
&=&\sum_{j_2,j_3,m_2,m_3}\Psi_{a^{\mu}23}^*X \Psi_{a^{\nu}23},\n
<a^{\mu}|X|a^{\nu}>_{np}
&=&\sum_{j_2,j_3,m_2,m_3}\Psi_{a^{\mu}32}^*X \Psi_{a^{\nu}23},\n
\Psi_{a^{\mu}23}&\equiv &
\sum_{j_1m_1}a^{\mu}_{j_1m_1}Tr Y_{j_3m_3} Y_{j_2m_2}Y_{j_1m_1}.
\label{avr1}
\eeqa
For simplicity, we restrict our consideration to
$U(1)$ gauge theory here
as it is straightforward to generalize it to $U(n)$ gauge theory.

Under the gauge transformation:
\beq
a_{\mu}\rightarrow P_{\mu}\Lambda+[a_{\mu},\Lambda ],
\eeq
the self-energy (\ref{fself}) changes as
\beqa
&&
{1\over 2}<\Lambda |tr\Gamma\cdot P{1\over \Gamma\cdot Q}\Gamma_{\nu}
{1\over \Gamma\cdot R}
|a^{\nu}>_p \n
&&
-{1\over 2}<a^{\mu}|tr\Gamma_{\mu}{1\over \Gamma\cdot Q}\Gamma\cdot P
{1\over \Gamma\cdot R}
|\Lambda>_p \n
&=&
-{1\over 2}<\Lambda |tr{1\over \Gamma\cdot Q}\Gamma_{\nu}|a^{\nu}>_p
-{1\over 2}<\Lambda |tr{1\over \Gamma\cdot R}\Gamma_{\nu}|a^{\nu}>_p\n
&&+{1\over 2}<a^{\nu}|tr{1\over \Gamma\cdot Q}\Gamma_{\nu}|\Lambda >_p
+{1\over 2}<a^{\nu}|tr{1\over \Gamma\cdot R}\Gamma_{\nu}|\Lambda >_p\n
&=&-8
\sum_{j_2,m_2}
Tr Y_{j_2m_2}^{\dagger} \Big({Q_{\nu}\over Q^2}Y_{j_2m_2}
\Big)[a^{\nu},\Lambda]\n
&&+8
\sum_{j_2,m_2}
Tr \Big({Q_{\nu}\over Q^2}Y_{j_2m_2}\Big)
Y_{j_2m_2}^{\dagger}[a^{\nu},\Lambda] ,
\label{gtr2p}
\eeqa
where we have retained only the linear terms in $a$.
The non-planar part is gauge invariant by itself
at the linearized level.

Since the fermionic contribution to the effective action
must be gauge invariant,
it is somewhat surprising to find that the gauge field self-energy is
not gauge invariant by itself.
The resolution of this puzzle is that the gauge invariance is restored
by the presence of the tadpole.
The tadpole (one point function) is
\beqa
&&-{1\over 2}
Tr\Gamma\cdot \delta A{1\over \Gamma\cdot Q}\n
&=&8\sum_{j_2,m_2}
Tr Y_{j_2m_2}^{\dagger} \Big({Q_{\mu}\over Q^2}Y_{j_2m_2}\Big)a^{\mu}\n
&&-8\sum_{j_2,m_2}
Tr \Big({Q_{\mu}\over Q^2}Y_{j_2m_2}\Big)Y_{j_2m_2}^{\dagger}a^{\mu} .
\label{tdplf}
\eeqa
The tadpole changes under gauge transformation as
\beqa
&&8\sum_{j_2,m_2}
Tr Y_{j_2m_2}^{\dagger} \Big({Q_{\mu}\over Q^2}Y_{j_2m_2}\Big)
[a^{\mu},\Lambda]\n
&&-8\sum_{j_2,m_2}
Tr \Big({Q_{\mu}\over Q^2}Y_{j_2m_2}\Big)Y_{j_2m_2}^{\dagger}
[a^{\mu},\Lambda] .
\label{gtr1p}
\eeqa
Since (\ref{gtr2p}) and (\ref{gtr1p}) cancel each other,
the fermionic contribution
to the one loop effective action is explicitly shown to be gauge
invariant at the linearized level.
The lesson we draw here is that the transversality of the gauge
field self-energy could be spoiled in NC gauge theory under the
presence of the tadpole.

We next investigate the gauge sector contribution.
In this section,
we deform the IIB matrix model action by adding a Myers term
in such a way that $S^2$ is a classical solution.
\beq
S_{IIB}+{i\over 3}f\epsilon_{\mu\nu\rho}Tr[A_{\mu}.A_{\nu}]A_{\rho} .
\label{defIIB}
\eeq
Since IIB matrix model is a large $N$ reduced model of
10d super Yang-Mills theory,
we also evaluate the gauge sector contributions in the
large $N$ reduced model of $D$ dimensional super Yang-Mills theory
with the identical deformation.

Firstly we evaluate the tadpole as:
\beq
(D-2)Tr\delta A \cdot Q{1\over Q^2} ,
\label{1point}
\eeq
where $D-1$ and $-1$ are the gauge field and ghost contributions
respectively.
It precisely cancels the fermionic contribution (\ref{tdplf})
when the physical degrees of freedom coincide
(in supersymmetric reduced models).

The gauge field sector contribution
to the planar part of the gauge field self-energy is
evaluated in Appendix A as
\beqa
&&{1\over 2}<a^{\mu}|{1\over Q^2R^2}
\Big(4P_{\mu}P_{\nu}-4P^2\delta_{\mu\nu}
-8\tilde{\delta}_{\mu\nu}
+8i\epsilon_{\mu\nu\rho}P^{\rho}
\n
&&
+(D-2)(Q_{\mu}R_{\nu}+Q_{\nu}R_{\mu}
-Q_{\nu}Q_{\mu}-R_{\nu}R_{\mu}
+(Q^2+R^2)\delta_{\mu\nu})\Big)
|a^{\nu}>_p ,
\label{gfself}
\eeqa
where $\tilde{\delta}_{\mu\nu}$ is Kronecker's $\delta$ in
the 3 dimensional sub-space in which $S^2$ extends.

Let us consider the $D=3$ case which is
relevant to a recent Monte-Carlo investigation\cite{ABNN}.
Firstly the one loop self-energy is of the same order with the tree action
if we fix $f^4N$ in the large $N$ limit.
It is because (\ref{gfself}) is $O(1/N)$ due to the $6j$ symbols
in the interaction vertices.
Secondly the self-energy is certainly positive for gauge fields
which carry small angular momentum.
In the case of
$n$ identical branes with $U(n)$ gauge group,
there are zero-modes $a_0^{\mu}$.
They represent the relative center of
mass positions of the branes.
For zero-modes, the gauge field self-energy can be estimated as
\beqa
&&{n^2\over 2}tr<a^{\mu}|{1\over Q^2R^2}
\Big({2\over 3}Q^2{\delta}_{\mu\nu}
-8{\delta}_{\mu\nu}
)|a^{\nu}>_p\n
&\sim&
{n^2\over 2N}
\Big({4\over 3}log(N/n)-8\Big)tra_0^{\mu}a_0^{\mu} ,
\label{3Dgfse}
\eeqa
where $tr$ is over the $n$ dimensional subspace of zero-modes.
We find that there arises positive mass term for zero-modes
in the large $N$ limit.
Therefore the degeneracy is lifted at the one loop level.
\footnote{With finite $N$,
it could become negative leading to an instability\cite{ABNN}.}

The fermionic contribution to the
planar part of the gauge field self-energy
(\ref{fself}) can be evaluated as
\beq
-{16\over 2}<a^{\mu}|{1\over Q^2R^2}\Big(
Q_{\mu}R_{\nu}+Q_{\nu}R_{\mu}-Q\cdot R\delta_{\mu\nu}
-2\tilde{\delta}_{\mu\nu}+\delta_{\mu\nu}
+i\epsilon_{\mu\nu\rho}P^{\rho}\Big)
|a^{\nu}>_p .
\label{fself2}
\eeq
After combining (\ref{gfself}) and (\ref{fself2}),
the planar part of the one loop self-energy of gauge field
in a deformed IIB matrix model is found as
\beq
{1\over 2}<a^{\mu}|{1\over Q^2R^2}
\Big(4P^2\delta_{\mu\nu}-4P_{\mu}P_{\nu}
+8(\tilde{\delta}_{\mu\nu}
-i\epsilon_{\mu\nu\rho}P^{\rho})
-16(\delta_{\mu\nu}-\tilde{\delta}_{\mu\nu})
\Big)|a^{\nu}>_p .
\label{quant}
\eeq

It is invariant under the linearized gauge transformation
$\delta a_{\mu}=P_{\mu}\Lambda$ just like
the self energy of gauge field in ordinary gauge theory.
However it contains mass and Chern-Simon type terms.
After inspecting (\ref{quant}), we note that
the one-loop self energy is positive for the
components of the gauge field
in the subspace in which $S^2$ extends.
However it is negative for the components in the extra-dimensions
as long as their angular momentum is small.
They are always interpreted as scalars from
gauge theory on $S^2$ point of view.
In the case of $n$ identical branes,
(\ref{quant}) is consistent with the one loop effective action
for zero-modes \cite{fuzS2}.

\section{Correlators on $S^2$}
\setcounter{equation}{0}

In this section, we evaluate the expectation values of
the following gauge invariant operator in NC gauge theory on $S^2$:
\beq
<{1\over N}TrA_{\mu}A_{\nu}> .
\label{invcor}
\eeq
Firstly we evaluate it through connected diagrams
expanding $A_{\mu}$ around a classical solution.
We then evaluate it through 1PI diagrams
after expanding $A_{\mu}$ around the quantum solution
which extremizes the effective action.
We find that the both methods give the identical
result as it can be argued to be the case.
Based on the explicit calculations up to the two loop level,
we identify the large $N$ scaling behavior of this
correlator. We further argue that this scaling behavior
is valid to all orders.

At the tree level, (\ref{invcor}) is evaluated as
\beq
{f^2\over N}Trp_{\mu}p_{\nu}
={f^2\over 3}l(l+1)\tilde{\delta}_{\mu\nu}
\sim {f^2N^2\over 12n^2}\tilde{\delta}_{\mu\nu} .
\label{tree}
\eeq
It is consistent with the fact that the space is
classically a fuzzy sphere of the radius $fl$.
In the classical limit this operator measures
the shape and extension of a space
realized in a matrix model.
It is hence a natural operator
to measure the geometry of the space
at the quantum level as well.

At the one loop level,
\beq
{f^2\over N}<Tra_{\mu}a_{\nu}>
={n^2\over f^2N}\sum_1^{2l}{2j+1\over j(j+1)}\delta_{\mu\nu}
\sim {2n^2\over f^2 N}log(N/n)\delta_{\mu\nu} .
\label{1loop}
\eeq
In a $D$ dimensional bosonic reduced model,
we also obtain the following tadpole
contribution due to the presence of the one point function (\ref{1point}):
\beq
{f^2\over N}<Tr p_{\mu}a_{\nu}+a_{\mu}p_{\nu}>
=-{D-2\over 3f^2}N\tilde{\delta}_{\mu\nu} .
\label{Dtdpl}
\eeq
It is of the same order with the tree contribution (\ref{tree})
if we fix $n^2/f^4N$ while (\ref{1loop}) is suppressed by
$log(N)/N^2$ in the large $N$ limit.
Thus the quantum effects tend to shrink the radius of the sphere.
This effect is caused by the logarithmic attractive
potential between eigenvalues in bosonic models.
In fact $n^2/f^4N$ cannot be too large since
otherwise (\ref{Dtdpl}) overwhelms (\ref{tree})
leading to unphysical results.

It is an indication of an instability of the sphere when the coupling
$n^2/f^4N$ is strong enough.
When $D=3$, the stability of a fuzzy sphere solution can be
estimated by the following one loop effective action 
with $<A_{\mu}>=\beta p_{\mu}$:
\beq
({1\over 2}\beta^4-{2\over 3}\beta^3f)Nl(l+1)
+N^2log\beta+N^2log(N/n) .
\eeq
The quantum vacuum is determine by minimizing it
with respect to $\beta$:
\beq
{1\over 2n^2}(\beta^3-\beta^2 f)N^3+N^2{1\over \beta}=0 .
\label{1lpeq}
\eeq

In the weak coupling limit $(\beta \rightarrow \infty)$,
this equation reproduces the classical solution $\beta=f$.
The deviation from the classical solution can be estimated from
(\ref{1lpeq})
to the leading order of ${n^2/f^4N}$ as
\beq
1 -{f\over \beta}=-2{n^2\over f^4N}+\cdots .
\eeq
The order parameter is estimated as
\beq
{\beta^2\over N}Trp_{\mu}p_{\nu}
={f^2N^2\over 12n^2}
\Big(1-4{n^2\over f^4N}+\cdots\Big)
\tilde{\delta}_{\mu\nu} ,
\eeq
which is consistent with (\ref{Dtdpl}).
We further note that the minimum of the effective action no longer exists when
\beq
{n^2\over f^4N}> {2^{9}\over 3^{3}}
\eeq
which indicates the instability of a fuzzy $S^2$ solution
beyond this coupling constant.
\footnote{This critical coupling was first estimated by
D. O'Connor.}
In fact this argument has been found to be in good agreement with a
Monte Carlo simulation\cite{ABNN}.

Here we return to a supersymmetric model:
the deformed IIB matrix model (\ref{defIIB}).
At the two loop level, we obtain from the one loop effective action
(\ref{quant})
\beqa
&&{f^2\over N}<Tra_{\mu}a_{\nu}>\n
&=&-{n^3\over f^6N}<{1\over (P^2)^2Q^2R^2}
\Big(4P^2\delta_{\mu\nu}-{4\over 3}P^2\tilde{\delta}_{\mu\nu}
+8\tilde{\delta}_{\mu\nu}
-16(\delta_{\mu\nu}-\tilde{\delta}_{\mu\nu})
\Big)>_p +~ n.p. ,\n
\label{aa}
\eeqa
where
\beqa
<X>_p
&=&\sum_{j_1,j_2,j_3,m_1,m_2,m_3}
\Psi^*_{123}X\Psi_{123},\n
\Psi_{123}&=&TrY_{j_3m_3}Y_{j_2m_2}Y_{j_1m_1} .
\eeqa
The symbol $n.p.$ denotes the non-planar contributions.
Each contribution in (\ref{aa}) is suppressed by $1/N^2$ compared to the
tree contribution when we fix $n^2/f^4N$ in the large $N$ limit.
Although there are no tadpoles at the one loop level
in supersymmetric models, they do arise
at the two loop level.
The total tadpole contribution
to this order parameter at the two loop level
is evaluated in Appendix B as
\beq
{f^2\over N}<Trp_{\mu}a_{\nu}+a_{\mu}p_{\nu}>
=-{n^3\over f^6N}\tilde{\delta}_{\mu\nu}{172\over 3}
<{1\over P^2Q^2R^2}>_p +~ n.p. .
\label{tdpl2l}
\eeq

The existence of the tadpole indicates the shift
of the classical solution due to the quantum effects.
Such a shift should also be calculable
by minimizing the two loop effective action
around $A_{\mu}=\beta p_{\mu}$.
We can read off the relevant effective action from
our previous results\cite{fuzS2}\cite{fuzS2S2} as
\beqa
&&({1\over 2}\beta^4-{2\over 3}\beta^3f){N^3\over 4n^2}\n
&-&{n^3\over \beta^4}\Big(32+4({f^2\over \beta^2})
-12(1-{f\over\beta})-32(1-{f\over\beta})
+24(1-{f\over\beta})\Big)
<{1\over P^2Q^2R^2}>_p +~ n.p. ,\n
\label{2lpact}
\eeqa
where the second line represents the two loop contribution.
The second term proportional to $({f^2/ \beta^2})$ denotes the
contribution from the Myers vertices.
The factor $(1-{f/\beta})$ represents the modification
of the gauge field propagator from the minimal one.
In this expression, we have retained only the linear
contributions with respect to this factor.
The third and fourth terms come from gauge and
fermionic contributions and
the last term comes from the Myers and cubic gauge vertices.
\footnote{Through this investigation, we have realized that the coefficient
12 in the third term of (\ref{2lpact})
was erroneously evaluated to be 10 in \cite{fuzS2S2}.
We have corrected this error in \cite{IT}.}

The deviation from the classical solution can be estimated
by minimizing (\ref{2lpact}) with respect to $\beta $ to the
leading order of $n^2/f^4N$ as
\beq
1-{f\over \beta}=-({n^5\over f^8 N^3})<{344\over P^2Q^2R^2}>_p +~ n.p. .
\eeq
We thus find that
\beq
{\beta^2\over N}Trp_{\mu}p_{\nu}
=(1-({n^5\over f^8 N^3})<{688\over P^2Q^2R^2}>_p \Big)
{f^2N^2\over 12n^2}\tilde{\delta}_{\mu\nu} +~ n.p.
\eeq
which agrees with (\ref{tdpl2l}).

Let us consider the large $N$ limit where
we fix $n^2/f^4N$ as the 't Hooft couplings.
They are appropriate 't Hooft
couplings in generic NC gauge theories on $S^2$.
It is because we obtain a factor $n/N$ at each
loop due to $6j$ symbols in the interaction vertices.
The other factor $n/f^4$
is generic in $U(n)$ gauge theory.
In such a large $N$ limit, the correlators scale as
follows to all orders
\beq
{1\over N}TrA_{\mu}A_{\nu} \sim N^{3\over 2}\tilde{\delta}_{\mu\nu} .
\label{S2scl}
\eeq
It is because the quantum corrections are small except the one loop
tadpole contributions in bosonic models as we have seen in this section.

We can explain this scaling behavior
by repeating the same argument with\cite{fuzS2S2}.
In this argument,
the large $N$ limit of the amplitudes are
estimated by their field theory counterparts.
It is because we obtain 2d gauge theory in a semiclassical limit.
In our case, the correlator is found to be logarithmically
divergent at the one loop level in (\ref{1loop}).
It corresponds to the logarithmic divergence of the
two point functions in field theory.
Since 2d gauge theory is super renormalizable,
higher loop corrections in 1PI diagrams are finite.

As for the tadpole contributions, they can be estimated
by the effective action.
The effective action is quadratically divergent at the one loop
level in bosonic models. It is the reason why the one loop tadpole
contribution  (\ref{Dtdpl}) is of the same
order with the tree estimates of the correlators in the
large $N$ limit. The higher order contributions are
suppressed by a power of $1/N^2$ due to super
renormalizability of 2d gauge theory.

With $U(1)$ gauge group in supersymmetric models,
we may be able to consider another scaling limit
where $1/f^4N^2$ is adopted as the 't Hooft coupling.
It is because another power of $1/N$ can be gained due to the cancellations
between the planar and non-planar amplitudes.
In this case, we find the following scaling behavior
\beq
{1\over N}TrA_{\mu}A_{\nu} \sim N\tilde{\delta}_{\mu\nu} ,
\label{S2scl}
\eeq
where it is again dominated by the tree contribution.
It is because the one loop corrections are $O(log(N))$
and the higher loop
corrections are $O(1)$ by the power counting arguments.
We therefore conclude that the order parameter on $S^2$
is always given by a unique tensor: Kronecker's $\delta$
in 3d in which $S^2$ extends. We note that
the behavior of the order parameter remains the same with the classical limit.

Another conclusion we can draw in this section is
that we only need to calculate 1PI diagrams
in our evaluation of the gauge invariant correlators.
In this procedure
we expand $A_{\mu}$ around
the quantum solutions of the effective action.
In fact we can give a generic argument to justify this
procedure as follows.

In order to calculate the correlators of $TrA_{\mu}A_{\nu}$,
it is sufficient to calculate the free energy
of the deformed action as
\beqa
e^{-F}&=&\int dAd\psi e^{-S-\eta^{\mu\nu}TrA_{\mu}A_{\nu}},\n
<TrA_{\mu}A_{\nu}>&=&{\delta F\over \delta \eta^{\mu\nu}}\Big|_
{\eta^{\mu\nu}=0} ,\cdots .
\eeqa
As it is well known, the free energy and the effective action are related by
the
Legendre transform:
\beqa
W[J]&=&\Gamma[\phi ]+J\cdot \phi,\n
J&=&{\delta \Gamma[\phi ]\over \delta \phi},
\eeqa
where $\phi$ denotes a generic field.
We can hence calculate $F=W[0]$ by minimizing the effective
action $\Gamma[\phi ]$.
The derivations with respect to $\eta^{\mu\nu}$
are equivalent to the insertions of $TrA_{\mu}A_{\nu}$
operator into the connected or 1PI diagrams respectively.

\section{Correlators on $S^2\times S^2$}
\setcounter{equation}{0}

In this section, we compute the vacuum expectation value of
the gauge invariant operators
$TrA_{\mu}A_{\nu}$ on a 4 dimensional homogeneous space: fuzzy $S^2\times S^2$.
The order parameters on such a space can be evaluated
in an analogous way as on $S^2$ in the preceding section.
However they exhibit a different scaling behavior
in the large $N$ limit due to different dimensionality.
We adopt the strategy to compute the 1PI diagrams
by expanding $A_{\mu}$ around a quantum
solution which extremizes the effective action
in this section.

The first option is to consider the following deformation:
\beq
S_{IIB}+{i\over 3}f_{\mu\nu\rho}Tr[A_{\mu},A_{\nu}]A_{\rho} ,
\eeq
where $f_{\mu\nu\rho}/f$ are the structure constants of
$SU(2)\times SU(2)$.
A fuzzy $S^2\times S^2$ is a classical solution of this model.
At the tree level, the order parameter is evaluated as
\beq
{f^2\over N}Trp_{\mu}p_{\nu}
={f^2\over 3}l(l+1)\breve{\delta}_{\mu\nu}
\sim {f^2N\over 12n}\breve{\delta}_{\mu\nu} ,
\label{tree2}
\eeq
where $N=n(2l+1)^2$.
We here assume that the both $S^2$ are of the identical size.
The symbol $\breve{\delta}_{\mu\nu}\equiv
f_{\mu\rho\sigma}f_{\nu\rho\sigma}/2f^2$
is Kronecker's $\delta$ in the 6 dimensional sub-space
in which $S^2\times S^2$ extends.
(\ref{tree2}) is consistent with the fact that the eigenvalues
of matrices are
distributed as $S^2\times S^2$ in the 6d space with the radius $fl$
at the classical level.

At the one loop level, we obtain
\beqa
{f^2\over N}<Tra_{\mu}a_{\nu}>
&=&{n^2\over f^2N}\sum_{j,p}{(2j+1)(2p+1)\over j(j+1)+p(p+1)}\delta_{\mu\nu}\n
&\sim& {n\over f^2 }2log(2)\delta_{\mu\nu} .
\label{1loop2}
\eeqa
Since we identify $n^2/f^4N$ as the 't Hooft coupling,
it is of the same order with the tree contribution (\ref{tree2}).

In order to evaluate the two loop contribution,
we first evaluate the
one loop self energy of gauge field.
This evaluation is reported in appendix C.
The order parameter $<TrA_{\mu}A_{\nu}>$ itself is quadratically divergent
by power counting
on 4 dimensional spaces like $S^2\times S^2$.
In this case, we can safely ignore non-planar contributions
since they are suppressed by a power of $N$.
The leading part of the gauge field self-energy (\ref{guge1lp}) is
\beq
{1\over 2}<a^{\mu}|{1\over Q^2R^2}
\Big(4P^2\delta_{\mu\nu}-4P_{\mu}P_{\nu}
\Big)|a^{\nu}>_p .
\label{quant2}
\eeq
The two loop contribution to the correlator
follows from this estimate as
\beqa
&&{f^2\over N}<Tra_{\mu}a_{\nu}>\n
&=&-{n^3\over f^6N}
(4\delta_{\mu\nu}-{2\over 3}\breve{\delta}_{\mu\nu})
<{1\over P^2Q^2R^2}>_p ,
\label{2loop2}
\eeqa
where
\beqa
<X>_p
&=&\sum_{j_i,p_i,m_i,q_i}
\Psi^*_{123}X\Psi_{123},\n
\Psi_{123}&=&TrY_{j_3m_3}Y_{j_2m_2}Y_{j_1m_1}
TrY_{p_3q_3}Y_{p_2q_2}Y_{p_1q_1} .
\eeqa

We also need to determine $<A_{\mu}>=\beta p_{\mu}$ by minimizing the
two loop effective action.
We can again read off the relevant effective action from
our previous results\cite{fuzS2S2}\cite{IT} as
\beqa
&&({1\over 2}\beta^4-{2\over 3}\beta^3f){N^2\over 2n}\n
&-&{n^3\over \beta^4}\Big(16F_4^p+8({f^2\over \beta^2})F_3^p
-12(1-{f\over\beta})F_3^p
+32(1-{f\over\beta})(F_3^p-F_4^p)
+24(1-{f\over\beta})F_3^p\Big) ,\n
\label{2lpact2}
\eeqa
where the first and
second line represent
the tree and two loop contributions respectively.
As in $S^2$ case: (\ref{2lpact}),
the second term proportional to $({f^2/ \beta^2})$ denotes the
contribution from the Myers vertices.
The third and fourth terms come from gauge and fermionic contributions and
the last term comes from the Myers and gauge cubic vertices.

We have ignored the one loop contribution $O(log(N))$
in comparison to the tree and the two loop contributions
of $O(N)$ in the large $N$ limit where $n^2/f^4N$ is fixed.
It is a natural 't Hooft coupling in 4d NC gauge theory
as well since
we obtain a factor of $n/N$ due to the $6j$ symbols
in the interaction vertices at each loop.
Since 4d gauge theory is renormalizable,
the over all degree of divergence remains the same to all orders.

The deviation from the classical solution can be estimated
by minimizing (\ref{2lpact2}) with respect to $\beta $ to the
leading order of $n^2/f^4N$ as
\beqa
1-{f\over \beta}
=-{n^4\over f^8N^2}(96F_4^p+4F_3^p) .
\label{4d2lp}
\eeqa
Since $F_3^p$ and $F_4^p$ are $O(1)$ in the large $N$ limit,
the two loop corrections are finite and small as long
as the 't Hooft coupling is small.
We have further checked that the two loop tadpole contributions are
consistent with this result in appendix C.

Alternatively we can evaluate the correlators on $S^2\times S^2$
in IIB matrix model itself.
In fact it has been found that a fuzzy $S^2\times S^2$
is a consistent solution of IIB matrix model
at the two loop level\cite{fuzS2S2}\cite{IT}.
Furthermore the most symmetric case where the both $S^2$
possess the identical radii is favored by the effective action.
It constitutes another evidence that 4 dimensionality
of space-time may be dynamically explained in IIB matrix model
\cite{BP}\cite{NS}\cite{Kyoto}.
In this case, $<A_{\mu}>=fp_{\mu}$ is
determined by minimizing the two loop effective action.
Expanding $A_{\mu}$ around the quantum solution,
the order parameter is given by the total of
(\ref{tree2}), (\ref{1loop2}) and (\ref{2loop2}):
\beqa
&&{f^2N\over 12n}\breve{\delta}_{\mu\nu}
+{n\over f^2 }2log(2)\delta_{\mu\nu}-{n^3\over f^6N}
(4\delta_{\mu\nu}-{2\over 3}\breve{\delta}_{\mu\nu})
F_3^p\n
&=&\sqrt{N}
({1\over 6\lambda}\breve{\delta}_{\mu\nu}
+\lambda log(2)\delta_{\mu\nu}
-\lambda^3{1\over 2}(\delta_{\mu\nu}-{1\over 6}\breve{\delta}_{\mu\nu})
F_3^p ) ,
\label{4dcor}
\eeqa
where $\lambda^2=4n^2/f^4N$.

However we have found $\lambda^2 =1/\sqrt{F^3_P}\sim 0.55$
at the extremum of the two loop effective action.
Therefore the two loop corrections in (\ref{4dcor})
are not small compared to the
tree and the one loop corrections. Since a naive perturbation theory is not
reliable,
we adopt a resummation procedure to use the one loop exact
propagator for gauge field.
In this resummation, we insert the one loop gauge field
self-energy into the gauge field propagator to all orders.

Let us investigate the one loop gauge field
self-energy in (\ref{quant2}) which is of the form
$(P^2\delta_{\mu\nu}-P_{\mu}P_{\nu})\omega (P)$:
\beqa
\omega (j_1,p_1)&=&{4\over f^4}\sum_{j_2,j_3,p_2,p_3}
{(2j_2+1)(2p_2+1)(2j_3+1)(2p_3+1)\over
(j_2(j_2+1)+p_2(p_2+1))(j_3(j_3+1)+p_3(p_3+1))}\n
&&\times     \left\{
 \begin{array}{ccc}
  j_1 &  j_2 & j_3 \\
  l   &  l & l
 \end{array}
\right\}^2
     \left\{
 \begin{array}{ccc}
  p_1 &  p_2 & p_3 \\
  l   &  l & l
 \end{array}
\right\}^2 ,
\eeqa
where we refer \cite{Edm} for $6j$ symbols.
In the large $N$ limit,  we can
adopt the Wigner approximation of
$6j$ symbols to estimate the above
amplitude\cite{IT}.
In such an approximation, it
reduces to the following integral
\beqa
\omega&=&{1\over f^4l^2\pi^2}\int_0^4 dYdZdMdN
{1\over (Y+M)(Z+N)}\n
&&\times{1\over \sqrt{2XY+2YZ+2ZX-X^2-Y^2-Z^2-XYZ}}\n
&&\times{1\over\sqrt{2LM+2MN+2NL-L^2-M^2-N^2-LMN}} ,
\eeqa
where $X=j_1^2/l^2,L=p_1^2/l^2$.
We can argue that this approximation is exact in the large $N$ limit
except at vanishingly small $X$ and $L$.
However such a region does not concern us since it does not contribute to the
quadratically divergent order parameter.

Integrating $Z$ and $M$, we obtain
\beqa
\omega&=&\lambda^2\int_0^4dYdN\n
&&\times{1\over \sqrt{X^2+Y^2+N^2+2XN+2YN-2XY-XYN}}\n
&&\times{1\over \sqrt{L^2+N^2+Y^2+2YL+2YN-2LN-YLN}} .
\eeqa
For small $X,L$, the self-energy logarithmically diverges:
\beq
\omega \sim \lambda^2log({4\over X+L})
\label{estslf}
\eeq
as in Figure 1 where we plot $\omega$ as a function of $X=L$.
A full $X,L$ dependence of
$\omega$ is plotted in Figure 2.

\begin{figure}[hbtp]
\epsfysize=8cm
\begin{center}
\vspace{1cm}
\hspace{0cm}
\epsfbox{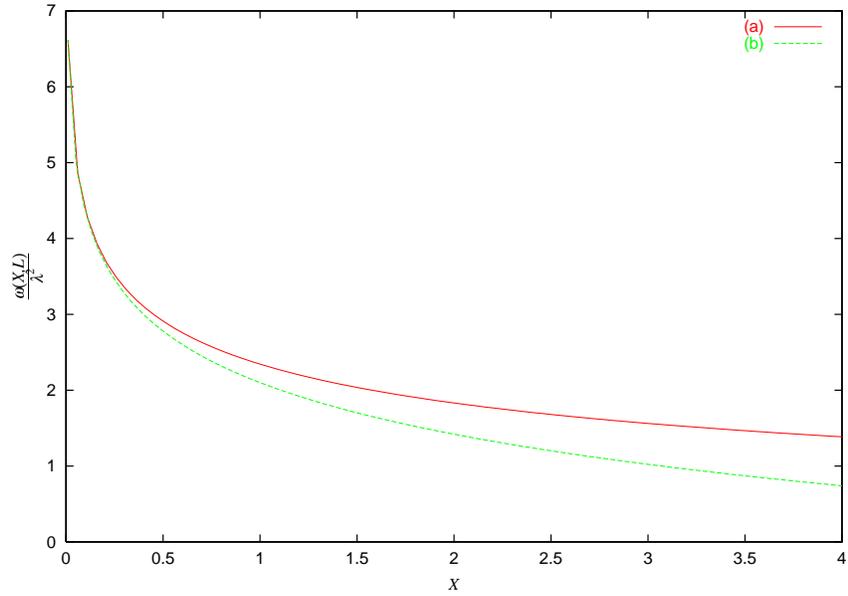}
\end{center}
\caption{
Plot of $\omega(X,L=X)$ against $X$:
\newline
Line (a) is  $\frac{\omega(X,L=X)}{\lambda^2}$
and line (b) is $0.98\ln(4/X)+0.74$.
}
\label{Fig:omega2d}
\end{figure}

\begin{figure}[hbtp]
\epsfysize=8cm
\begin{center}
\vspace{1cm}
\hspace{0cm}
\epsfbox{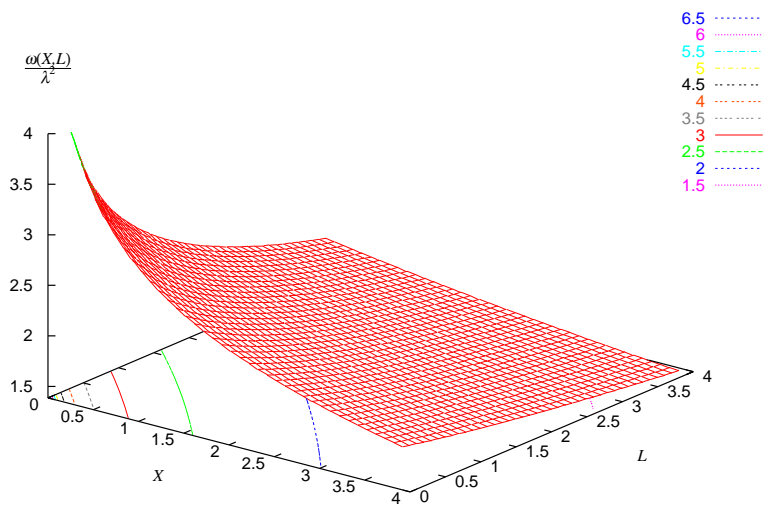}
\end{center}
\caption{Plot of $\omega(X,L)$ against $X$ and $L$}
\label{Fig:omega3d}
\end{figure}

In terms of $\omega$,
the  one loop exact propagator
to the leading order in $1/P^2$ expansion is
\beq
{1\over P^2}{1\over 1+\omega}\delta_{\mu\nu}
+{P_{\mu}P_{\nu}\over P^4}{\omega\over 1+\omega} .
\label{1pexpr}
\eeq
The order parameter is given in this resummation scheme as
\beqa
&&\sqrt{N}
{1\over 6\lambda}\breve{\delta}_{\mu\nu}+{f^2\over N}<Tra_{\mu}a_{\nu}>\n
&=&\sqrt{N}{1\over 6\lambda}\breve{\delta}_{\mu\nu}
+{1\over f^2N}\sum_{j,p}{(2j+1)(2p+1)\over j(j+1)+p(p+1)}{1\over 1+\omega
(j,p)}
\delta^{\mu\nu}\n
&&+{1\over 6f^2N}\sum_{j,p}{(2j+1)(2p+1)\over j(j+1)+p(p+1)}
{\omega (j,p)\over 1+\omega (j,p)} \breve{\delta}_{\mu\nu} .\n
\eeqa
In our approximation, it reduces to the following integrals
\beqa
&&\sqrt{N}{1\over 6\lambda}\breve{\delta}_{\mu\nu} +
\sqrt{N}{\lambda\over 2} \int_0^4 dXdL{1\over
X+L} {1\over 1+\omega (X,L)}\delta_{\mu\nu}\n
&&+\sqrt{N}{\lambda\over 12} \int_0^4 dXdL{1\over
X+L} {\omega (X,L)\over 1+\omega (X,L)}\breve{\delta}_{\mu\nu} .
\label{1lpexcr}
\eeqa
We can numerically
estimate the above integrals with $\lambda^2\sim 0.55$ as
\beq
\sqrt{N}\Big(0.48\breve{\delta}_{\mu\nu} + 0.91\delta_{\mu\nu}\Big) .
\label{fincor}
\eeq

Although this estimate is not exact,
it is a universal conclusion that the gauge invariant correlator
$<{1\over N}TrA_{\mu}A_{\nu}>$ scales as $\sqrt{N}$ in
4 dimensional homogeneous spaces $S^2\times S^2$
in the large $N$ limit.  It is because
we can estimate the scaling behavior of the correlators
to all orders by the power counting arguments.
In 4d, gauge theory is renormalizable. Therefore the
over all degree of the divergence of the correlators
remains the same to all orders.
The scaling behavior of the order parameter
$(\sqrt{N})$ is the reflection of the fact that it
is quadratically divergent. Hence this scaling behavior is valid to all orders
with our identification of the
't Hooft couplings.

The order parameter consists of the two independent
tensors $\breve{\delta}_{\mu\nu}$ and $\delta_{\mu\nu}$.
The first part can be interpreted as a distribution of
$S^2\times S^2$ extending in the 6 dimensional space
just like our background.
Since it is $O(R^2)$
in term of its radius $R$, we may identify $R\sim N^{1/4}$
from (\ref{fincor}).
The non-commutativity scale (volume of a quantum) is set by the
't Hooft couplings.

The second
term proportional to $\delta_{\mu\nu}$
solely arises from quantum fluctuations.
The identical $N$ dependence indicates that it also represents
a 4 dimensional distribution.
However $\delta_{\mu\nu}$ implies that it extends in
10 dimensions.
We interpret it as the contributions from a fractal whose
fractal dimension is 4. A concrete example
is the branched polymers which have
appeared in a previous study of IIB matrix
model around commutative backgrounds\cite{BP}.
Furthermore we have argued that 4d fuzzy homogeneous spaces
are smoothly connected with the branched polymers\cite{fuzS2S2}.
The large $N$ scaling behavior of the gauge invariant correlator (\ref{fincor})
is  consistent with our viewpoint.

\section{Connected two point functions}
\setcounter{equation}{0}

In this section, we investigate the connected two point functions
of the gauge invariant operators ${1\over N}TrA_{\mu}A_{\nu}$.
In particular we are interested in the operators ($\mu\neq\nu$)
with no vacuum expectation values.
Let us consider the connected two point
functions of them on $S^2$.
It is most convenient to expand the functions (or matrices)
in terms of spherical harmonics $Y_{lm}$.
Since $p_{\mu}$ can be identified with $Y_{1\mu}$,
this operator can be regarded as the gauge invariant completion
of Tr$(Y_{l\mu}a_{\nu}+Y_{l\nu}a_{\mu})$, namely
a Wilson line on $S^2$ which carries the minimum angular
momentum\cite{IIKK}\cite{Gross}.

At the tree level, the connected two point functions
can be evaluated as
\beqa
&&<{f^2\over N}Tr(p_{\mu}a_{\nu}+a_{\mu}p_{\nu})
{f^2\over N}Tr(p_{\rho}a_{\sigma}+a_{\rho}p_{\sigma})>\n
&=&{N\over 12n^2}
\Big(\delta_{\mu\rho}\delta_{\nu\sigma} +
\delta_{\mu\sigma}\delta_{\nu\rho}
\Big) .
\label{tree2p}
\eeqa
At the one loop level, there occurs the following
contribution
\beqa
&&<{f^2\over N}Tra_{\mu}a_{\nu}
{f^2\over N}Tra_{\rho}a_{\sigma}>\n
&=& {n^2\over f^4N^2}\sum_j{2j+1\over (j(j+1))^2}
\Big(\delta_{\mu\rho}\delta_{\nu\sigma} +
\delta_{\mu\sigma}\delta_{\nu\rho}
\Big)\n
&=&{n^2\over f^4N^2}
\Big(\delta_{\mu\rho}\delta_{\nu\sigma} +
\delta_{\mu\sigma}\delta_{\nu\rho}
\Big) .
\eeqa
This contribution vanishes in comparison to the tree contribution
(\ref{tree2p}) in the large $N$ limit where
either $n^2/f^4N$ or $1/f^4N^2$ is fixed.
We do find nonvanishing contributions
to this correlator at the one loop level
through the quantum corrections of
the gauge field propagator.
In the case of the 3D bosonic matrix model,
the one loop self-energy of gauge field with small
angular momentum scales as in (\ref{3Dgfse}).
With the self-energy insertion into the
propagator, the two point function behaves as
\beq
-{N\over 12n^2}
\Big(\delta_{\mu\rho}\delta_{\nu\sigma} +
\delta_{\mu\sigma}\delta_{\nu\rho}
\Big)
{{2\over 3}\tilde{\lambda}^2 log(N/n)} ,
\eeq
where $\tilde{\lambda}^2=n^2/f^4N$.
We thus find that the two point functions $NG_2$ on $S^2$
exhibits the logarithmic scaling violation
at the one loop level.
Since we can trace it to the logarithmically divergent
one loop mass term,
these logarithms may be resummed as
\beq
G_2\sim {1\over 2+{4\over 3}\tilde{\lambda}^2 log(N/n)}
{1\over 6n^2}
\Big(\delta_{\mu\rho}\delta_{\nu\sigma} +
\delta_{\mu\sigma}\delta_{\nu\rho}
\Big) .
\eeq
In the case of supersymmetric models, the corresponding logarithmic
scaling violations are absent since
quantum corrections are finite to all orders in the large $N$ limit.

We next investigate the connected two point functions
of these Wilson lines on $S^2\times S^2$
in IIB matrix model.
At the tree level,
\beqa
&&<{f^2\over N}Tr(p_{\mu}a_{\nu}+a_{\mu}p_{\nu})
{f^2\over N}Tr(p_{\rho}a_{\sigma}+a_{\rho}p_{\sigma})>\n
&=&{1\over 12n}
\Big(\delta_{\mu\rho}\delta_{\nu\sigma} +
\delta_{\mu\sigma}\delta_{\nu\rho}
\Big) .
\label{tree4d}
\eeqa
At the one loop level, there occurs the following
contribution
\beqa
&&<{f^2\over N}Tra_{\mu}a_{\nu}
{f^2\over N}Tra_{\rho}a_{\sigma}>\n
&=& {n^2\over f^4N^2}\sum_{j,p}{(2j+1)(2p+1)\over (j(j+1)+p(p+1))^2}
\Big(\delta_{\mu\rho}\delta_{\nu\sigma} +
\delta_{\mu\sigma}\delta_{\nu\rho}
\Big)\n
&=&{n^2\over f^4N^2}log(N/n)
\Big(\delta_{\mu\rho}\delta_{\nu\sigma} +
\delta_{\mu\sigma}\delta_{\nu\rho}
\Big) .
\eeqa
This term is suppressed by $1/N$ in comparison to the tree term
(\ref{tree4d}) in the large $N$ limit where
$n^2/f^4N$ is fixed.
Therefore the nontrivial contributions in the large
$N$ limit also come from the quantum corrections of the
gauge field propagator on $S^2\times S^2$.

With the one loop self-energy estimation (\ref{estslf}),
we obtain
\beqa
&& -{1\over 12n}{\lambda^2log(N/n)}
\Big(\delta_{\mu\rho}\delta_{\nu\sigma} +
\delta_{\mu\sigma}\delta_{\nu\rho}
\Big) .
\eeqa
We thus find the logarithmic scaling violation
of the two point functions $G_4$ on $S^2\times S^2$ also.
We can trace them to the wave function renormalization
of gauge field. It is therefore likely that
these logarithms may be summed by the renormalization group.
Let us contemplate to renormalize the gauge field as
\beqa
A_{\mu}&=&p_{\mu}+Z_{j,p}a_{\mu}^{j,p}Y_j\otimes Y_p,\n
Z_{j,p}^2&=&1-\lambda^2log\Big((N/n)(j^2+p^2)\Big) .
\eeqa
It is then natural to renormalize the two point functions of the Wilson lines
as
\beq
G_4^R=Z^{-2}G_4 .
\eeq
By assuming $G_4^R$ possesses a finite
large $N$ limit, we obtain the following renormalization group
equation for bare $G_4$
\beq
N{\partial\over \partial N}G_4=-\lambda^2G_4 .
\eeq
This equation suggests that $G_4$ scales as
$(N/n)^{-\lambda^2}$ in the large $N$ limit with $\lambda^2$ being fixed.

It is clear that we need to renormalize NC gauge theory on homogeneous
spaces to determine the scaling behavior of the Wilson lines.
Admittedly our investigation in this section
is exploratory and a more detailed investigation
on this problem is necessary.
We also need to study the effects of the non-planar contributions.

\section{Conclusions and Discussions}
\setcounter{equation}{0}

In this paper, we have investigated correlators of
$TrA_{\mu}A_{\nu}$ in matrix models of homogeneous spaces
such as $S^2$ and $S^2\times S^2$.
It measures the geometry of Euclidean space-time
on which NC gauge theory is realized.
Since space-time and matter are unified in matrix models,
the structure of space-time in general receives quantum corrections.
This feature also suggests a close connection between
NC gauge theory and quantum gravity.

We have investigated these correlators
in matrix models up to two loop level.
We have developed a procedure to evaluate
them through 1PI diagrams.
We have determined the large $N$ scaling behavior
of the geometric order parameters to all orders.

On 2 dimensional spaces $S^2$, we have found that
the expectation values of them are given by a unique tensor
just like in the classical limit.
This fact shows that the eigenvalues of matrices
are always distributed as $S^2$.
On 4 dimensional spaces $S^2\times S^2$, we have found that
these order parameters consist of two independent tensors.
Although the first tensor is due to a distribution
of $S^2\times S^2$ , the second tensor
indicates a distributions of a 4d fractal.
We conclude that fuzzy $S^2\times S^2$
acquires a 4 dimensional fractal structure
due to NC gauge theory on it
in contrast to fuzzy $S^2$.

The operators with vanishing vacuum expectation
values are the Wilson lines which carry the minimum momentum.
We have investigated the two point functions of these Wilson lines
on $S^2$ and $S^2\times S^2$.
We have found logarithmic scaling violations
(or finite anomalous dimensions) of the
Wilson lines.
We note that
it is a suppression effect for low multipoles
which might be relevant to such a tendency observed
in the cosmic microwave background.
We believe that our procedure developed in this paper should be useful
to study more generic Wilson lines.
We hope that such investigations will further
elucidate gravitational aspects of NC gauge theory.

In the context of IIB matrix model, fuzzy $S^2\times S^2$
solutions may be interpreted as metastable D3-branes.
We are most interested in the ultimate ground state
into which these branes eventually decay.
This is an analogue of tachyon condensation problem
of unstable D-branes in IIB matrix model\cite{Sen}.
We also hope to make progress toward understanding
this fundamental question.

\begin{center} \begin{large}
Acknowledgments
\end{large} \end{center}
This work is supported in part by the Grant-in-Aid for Scientific
Research from the Ministry of Education, Science and Culture of Japan.

\section*{Appendix A}
\renewcommand{\theequation}{A.\arabic{equation}}
\setcounter{equation}{0}

In this appendix, we evaluate the one loop gauge field self-energy
in $D$ dimensional bosonic reduced models with Myers term.
We list the contributions from
different diagrams separately. We only show
planar contributions since non-planar contributions
can be obtained from them as in (\ref{avr1})

\begin{itemize}
\item From cubic gauge couplings:
\beqa
&&{1\over 2}<a^{\mu}|{1\over Q^2R^2}
\Big(4P_{\nu}P_{\mu}
-(D-2)(Q_{\nu}Q_{\mu}+R_{\nu}R_{\mu})\n
&&
+(D-1)(Q_{\mu}R_{\nu}+Q_{\nu}R_{\mu})
-(4P^2+Q^2+R^2)\delta_{\mu\nu}\Big)
|a^{\nu}>_p .\n
\eeqa

\item From quartic gauge coupling:
\beq
{D-1\over 2}<a^{\mu}|{1\over
Q^2R^2}\Big((Q^2+R^2)\delta_{\mu\nu}\Big)|a^{\nu}>_p .
\eeq

\item From ghost:
\beq
-{1\over 2}<a^{\mu}|{1\over
Q^2R^2}\Big(Q_{\mu}R_{\nu}+Q_{\nu}R_{\mu}\Big)|a^{\nu}>_p .
\eeq

\item From Myers terms:
\beq
-{8\over 2}<a^{\mu}|{1\over Q^2R^2}\tilde{\delta}_{\mu\nu}|a^{\nu}>_p .
\eeq
where $\tilde{\delta}_{\mu\nu}\equiv
\epsilon_{\mu\rho\sigma}\epsilon_{\nu\rho\sigma}/2$ is the projector into
the 3 dimensional sub-space in which $S^2$ extends.

\item From Myers term and cubic gauge coupling:
\beq
{12\over 2}<a^{\mu}|{1\over Q^2R^2}i\epsilon_{\mu\nu\rho}P^{\rho}|a^{\nu}>_p .
\eeq

\end{itemize}

In total we obtain (\ref{gfself}) in section 2.

We further list the planar part of the
ghost and fermion self-energy at the one loop level for completeness.

\begin{itemize}

\item
The ghost self energy:
\beq
-<b|{1\over Q^2R^2}P^2|c>_p  .
\eeq

\item
The fermion self energy:
\beqa
&&-4<\bar{\psi}|{1\over Q^2R^2}P\cdot\Gamma |\psi>_p\n
&&-i<\bar{\psi}|{1\over (Q^2)^2R^2}\epsilon_{\mu\nu\rho}
\Gamma^{\mu\nu\sigma}Q^{\rho}Q^{\sigma} |\psi>_p
-i<\bar{\psi}|{1\over Q^2(R^2)^2}\epsilon_{\mu\nu\rho}
\Gamma^{\mu\nu\sigma}R^{\rho}R^{\sigma} |\psi>_p . \n
&&
\eeqa

\end{itemize}

\section*{Appendix B}
\renewcommand{\theequation}{B.\arabic{equation}}
\setcounter{equation}{0}

In this appendix, we evaluate the
tadpole contributions to the following
correlator on $S^2$:
\beq
{f^2\over N}<Trp_{\mu}a_{\nu}+a_{\mu}p_{\nu}> .
\label{appa2}
\eeq

The relevant cubic vertices in a deformed IIB matrix model are
\beq
f^4Tr\Big(P_{\mu}a_{\nu}[a_{\mu},a_{\nu}]
-{i\over 3}\epsilon_{\mu\nu\rho}a_{\mu}[a_{\nu},a_{\rho}]
+{1\over 2}\bar{\psi}\Gamma_{\mu}[a^{\mu},\psi]
+P_{\mu}b[a_{\mu},c] \Big).
\eeq
By attaching the one loop self-energy of gauge field (\ref{quant})
to the cubic vertices, we evaluate the pure gauge field contribution
to (\ref{appa2}) as
\beq
{n^3\over f^6N}\tilde{\delta}_{\mu\nu}
\Big(<{12\over Q^2R^2}>_p-{4\over 3}<{1\over P^2Q^2R^2}>_p
-{104\over 3}<{1\over P^2Q^2R^2}>_p
\Big)+~ n.p. .
\label{ap}
\eeq
We note that the contributions from the longitudinal part of
the gauge field self-energy (the second terms in
(\ref{aa}) and (\ref{ap})) to the gauge invariant
correlator cancel out each other.

The remaining contributions are the followings.
\begin{itemize}
\item
The ghost contribution
\beq
{n^3\over f^6N}\tilde{\delta}_{\mu\nu}
{1\over 3}<{1\over Q^2R^2}>_p +~ n.p. .
\eeq

\item
The fermion contribution
\beq
{n^3\over f^6N}\tilde{\delta}_{\mu\nu}
\Big(-{64\over 3}<{1\over Q^2R^2}>_p
-{64\over 3}<{1\over P^2Q^2R^2}>_p
\Big) +~ n.p. .
\eeq

\item
The quartic gauge coupling contribution
\beq
{n^3\over f^6N}\tilde{\delta}_{\mu\nu}
<{9\over Q^2R^2}>_p +~ n.p. .
\eeq

In total we obtain (\ref{tdpl2l}) in section 3.

\end{itemize}

\section*{Appendix C}
\renewcommand{\theequation}{C.\arabic{equation}}
\setcounter{equation}{0}

In this appendix, we evaluate
the planar part of the one loop gauge field self-energy
on $S^2\times S^2$.
We subsequently
carry out a perturbative evaluation of
the tadpole contributions to the following
order parameter on $S^2\times S^2$:
\begin{eqnarray}
\frac{f^2}{N}<Tr p_{\mu}a_{\nu}+a_{\mu}p_{\nu}>\qquad
\;.
\label{tad}
\end{eqnarray}

The contribution to the gauge field self-energy
from the gauge sector is
\beqa
&&{1\over 2}<a^{\mu}|{1\over Q^2R^2}
\Big(4P_{\mu}P_{\nu}-4P^2\delta_{\mu\nu}
-8\breve{\delta}_{\mu\nu}
+8if_{\mu\nu\rho}P^{\rho}
\n
&&
+(D-2)(Q_{\mu}R_{\nu}+Q_{\nu}R_{\mu}
-Q_{\nu}Q_{\mu}-R_{\nu}R_{\mu}
+(Q^2+R^2)\delta_{\mu\nu})\Big)
|a^{\nu}>_p ,
\label{4dgf}
\eeqa
where in this context
\beqa
<a^{\mu}|X|a^{\nu}>_p
&=&\sum_{j_i,p_i,m_i,q_i}
\Psi^*_{a^{\mu}23}X\Psi_{a^{\nu}23},\n
\Psi_{a^{\mu}23}&=&\sum_{j_1m_1p_1q_1}
a^{\mu}_{\j_1m_1p_1q_1}TrY_{j_3m_3}Y_{j_2m_2}Y_{j_1m_1}
TrY_{p_3q_3}Y_{p_2q_2}Y_{p_1q_1} .
\eeqa

The fermion sector contribution has the form as the following
\begin{eqnarray}
&&\frac{16}{2}<a_{\mu}|
\frac{n}{P^2R^2}\Bigg[
Q_{\mu}R_{\nu+}Q_{\nu}R_{\mu}-Q\cdot R\;\delta_{\mu\nu} \nonumber\\
&&-i\bar{f}_{\mu\nu\rho}\left(
\frac{\bar{Q}_{\rho}(\tilde{Q}\cdot\tilde{R})+\bar{R}_{\rho}\bar{Q}^2}{Q^2}
+\frac{\bar{Q}_{\rho}\bar{R}^2+\bar{R}_{\rho}(\tilde{Q}\cdot\tilde{R})}{R^2}
\right) \nonumber \\
 &&-i\tilde{f}_{\mu\nu\rho}\left(
\frac{\tilde{Q}_{\rho}(\bar{Q}\cdot\bar{R})+\tilde{R}_{\rho}\tilde{Q}^2}{Q^2}
+\frac{\tilde{Q}_{\rho}\tilde{R}^2+\tilde{R}_{\rho}(\bar{Q}\cdot\bar{R})}{R^2}
\right) \nonumber \\
&&-\frac{1}{Q^2R^2}\Big(
(\bar{Q}^2\bar{R}^2-\tilde{Q}^2\tilde{R}^2)\bar{\delta}_{\mu\nu}
+(\tilde{Q}^2\tilde{R}^2-\bar{Q}^2\bar{R}^2)\tilde{\delta}_{\mu\nu}
\nonumber \\
&&\bar{Q}_{\mu}\tilde{Q}_{\nu}+\bar{R}_{\mu}\tilde{R}_{\nu}
-\delta_{ij}(\bar{Q}^2\bar{R}^2+\tilde{Q}^2\tilde{R}^2+2(\bar{Q}\cdot\bar{R})
(\tilde{Q}\cdot\tilde{R})
)
\Big)
\Bigg]  |a_{\nu}>_{p} ,\nonumber  \\
\label{fermi}
\end{eqnarray}
where $\bar{Q}_{\mu}, \bar{R}_{\mu}$ and $\tilde{Q}_{\mu}, \tilde{R}_{\mu}$
denote the components of the operators in the first and second three
dimensional subspaces.
$\delta_{ij}$ is the Kronecker's $\delta$ in the remaining dimensions.

We restrict our attention in the case of D=10 for simplicity.
Using (\ref{fermi}) we obtain the 1-loop  gauge field self-energy on
$S^2\times S^2$:
\begin{eqnarray}
&&\frac{1}{2}<a_{\mu}|\frac{n}{Q^2R^2}\Bigg[
4P_{\mu}P_{\nu}-4P^2\delta_{\mu\nu}
-8\breve{\delta}_{\mu\nu}+8if_{\mu\nu\rho}P^{\rho}\n
&&-16i\bar{f}_{\mu\nu\rho}\left(
\frac{\bar{Q}_{\rho}(\tilde{Q}\cdot\tilde{R})+\bar{R}_{\rho}\bar{Q}^2}{Q^2}
+\frac{\bar{Q}_{\rho}\bar{R}^2+\bar{R}_{\rho}(\tilde{Q}\cdot\tilde{R})}{R^2}
\right) \nonumber \\
 &&-16i\tilde{f}_{\mu\nu\rho}\left(
\frac{\tilde{Q}_{\rho}(\bar{Q}\cdot\bar{R})+\tilde{R}_{\rho}\tilde{Q}^2}{Q^2}
+\frac{\tilde{Q}_{\rho}\tilde{R}^2+\tilde{R}_{\rho}(\bar{Q}\cdot\bar{R})}{R^2}
\right) \nonumber \\
&&-\frac{16}{Q^2R^2}\Big(
(\bar{Q}^2\bar{R}^2-\tilde{Q}^2\tilde{R}^2)\bar{\delta}_{\mu\nu}
+(\tilde{Q}^2\tilde{R}^2-\bar{Q}^2\bar{R}^2)\tilde{\delta}_{\mu\nu}
\nonumber \\
&&\bar{Q}_{\mu}\tilde{Q}_{\nu}+\bar{R}_{\mu}\tilde{R}_{\nu}
-\delta_{ij}(\bar{Q}^2\bar{R}^2+\tilde{Q}^2\tilde{R}^2+2(\bar{Q}\cdot\bar{R})
(\tilde{Q}\cdot\tilde{R})
)
\Big)\Bigg]
|a_{\nu}>_p .
\label{guge1lp}
\end{eqnarray}

Now we can evaluate the order parameter (\ref{tad}) at the 2-loop level.
There are contributions from  cubic gauge,  ghost, fermion, quartic gauge
couplings
which correspond to (B.3)-(B.6).
In total we obtain
\begin{eqnarray}
\frac{n^3}{f^6N}\breve{\delta}_{\mu\nu}\left(
-12F_{3}^p-\frac{32}{3}F_{4}^p+\frac{64}{3}G^p \right)\;,
\label{tad2}
\end{eqnarray}
where
\begin{eqnarray}
F_{4}^p&=&2\left<\frac{\bar{Q}^2\bar{R}^2+\tilde{Q}^2\tilde{R}^2
+2(\bar{Q}\cdot\bar{R})(\tilde{Q}\cdot\tilde{R})}{P^2Q^4R^4}\right>_p\;,
\label{F4p}\n
G^p&=&\left<\frac{1}{P^2Q^4R^2}\Big[
(\bar{P}\cdot\bar{R})\bar{Q}^2+(\bar{P}\cdot\bar{Q})(\tilde{Q}\cdot\tilde{R})
+(\tilde{P}\cdot\tilde{R})\tilde{Q}^2+(\tilde{P}\cdot\tilde{Q})(\bar{Q}\cdot
\bar{R})
\Big]\right>_p
\nonumber \\
&=&\frac{1}{2}F_{3}^p-\frac{1}{2}F_{4}^p\;.
\label{Gp}
\end{eqnarray}
Finally we obtain
\begin{eqnarray}
\frac{f^2}{N}<Tr p_{\mu}a_{\nu}+a_{\mu}p_{\nu}>&=&
-\frac{n^3}{f^6N}\frac{\breve{\delta}_{\mu\nu}
}{3}(4F_{3}^p+96F_{4}^p)\;.
\label{tad3}
\end{eqnarray}
The result (\ref{tad3}) is indeed consistent with (\ref{4d2lp}).

\end{document}